\documentclass[12pt,a4paper]{article}
\usepackage{graphics}
\usepackage{color}
\usepackage{bbm}
\usepackage{amsmath}

\newcommand{\bfm}[1]{\textbf{\em #1}}
\newcommand{\bfs}[1]{\boldsymbol{#1}}

\newcommand{\fem}[1]{{\rm\bf #1}}

\newcommand\const{\mbox{const}}

\DeclareMathOperator{\tr}{tr}
\DeclareMathOperator{\dev}{dev}

\graphicspath{{../Fig/}{../Fig_FiniteStrain/}{../Fig_Xfig/}}

\newcommand{\inclps}[3]{\resizebox{#1}{#2}{\includegraphics{#3}}}

\begin{document}

\title{Finite-strain formulation and FE implementation of~a~constitutive model
       for powder compaction}
\author{S. Stupkiewicz$^{a,b,}$\footnote{Corresponding author}, A. Piccolroaz$^{b}$, D. Bigoni$^{b}$\\[1ex]
\normalsize{${}^{a}$ Institute of Fundamental Technological Research (IPPT),} \\[-0.5ex]
\normalsize{Pawi\'{n}skiego 5b, 02-106 Warsaw, Poland} \\
\normalsize{${}^{b}$ University of Trento, via Mesiano 77, I-38123 Trento, Italy} \\
\normalsize{e-mail: sstupkie@ippt.pan.pl; roaz@ing.unitn.it; bigoni@unitn.it} }

\date{}

\maketitle

\begin{abstract}
A finite-strain formulation is developed, implemented and tested for a constitutive
model capable of describing the transition from granular to fully dense state during
cold forming of ceramic powder. This constitutive model (as well as many others
employed for geomaterials) embodies a number of features, such as pressure-sensitive
yielding, complex hardening rules and elastoplastic coupling, posing considerable
problems in a finite-strain formulation and numerical implementation. A number of
strategies are proposed to overcome the related problems, in particular, a
neo-Hookean type of modification to the elastic potential and the adoption of the
second Piola-Kirchhoff stress referred to the intermediate configuration to describe
yielding. An incremental scheme compatible with the formulation for elastoplastic
coupling at finite strain is also developed, and the corresponding constitutive
update problem is solved by applying a return mapping algorithm.
\end{abstract}


\section{Introduction}

The formulation and implementation of elastoplastic constitutive equations for
metals at large strain have been thoroughly analyzed in the last thirty years, see
for instance \cite{SimoHughes98,SouzaNeto08}, so that nowadays they follow accepted
strategies. For these materials, pressure-insensitive yielding, $J_3$-independence,
and incompressibility of plastic flow strongly simplify the mechanical behaviour,
while frictional-cohesive and rock-like materials (such as granular media, soils,
concretes, rocks, ceramics and powders) are characterized by pressure-sensitive,
$J_3$-dependent yielding, dilatant/contractant flow, nonlinear elastic behaviour
even at small strain and elastoplastic coupling. There have been several attempts to
generalize treatment of metal plasticity at large strain in this context
\cite{Borja98,MeschkeLiu99,Perez03,Ortiz04,Rouainia06,Frenning07,Karrech11}, but
many problems still remain not completely solved. These include the form of the
elastic potential, the stress measure to be employed in the yield function, which
has to provide an easy interpretation of experiments, the flow rule and the
elastic-plastic coupling laws.

The main difficulty in the practical application of finite-strain elastoplasticity
models, as opposed to their small-strain counterparts, is related to development and
implementation of incremental (i.e., finite-step) constitutive relationships. The
difficulties lie, for instance, in formulation and solution of the highly nonlinear
constitutive update problem, consistent treatment of plastic incompressibility (or
plastic volume changes), and consistent linearization of the incremental
relationships. The last issue is of the utmost importance for overall computational
efficiency of the finite element models because consistent linearization (consistent
tangent) is needed to achieve the quadratic convergence of the Newton method.

\begin{figure}[!htcb]
  \centerline{
    \begin{tabular}{ccc}
      \inclps{0.4\textwidth}{!}{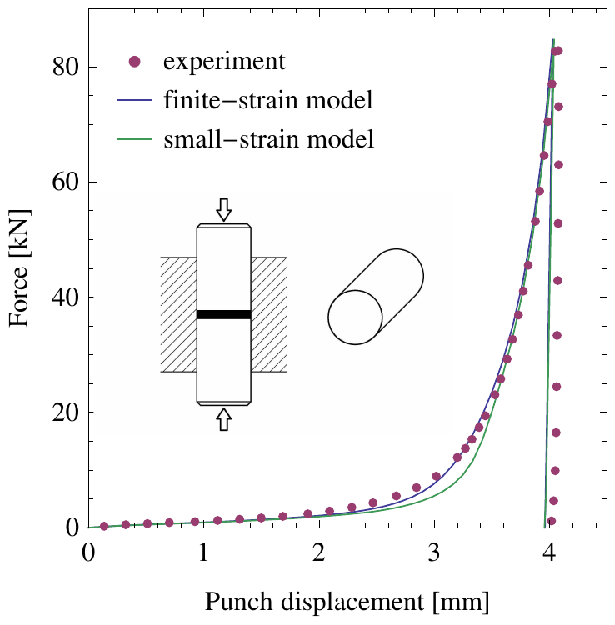} & &
      \inclps{0.4\textwidth}{!}{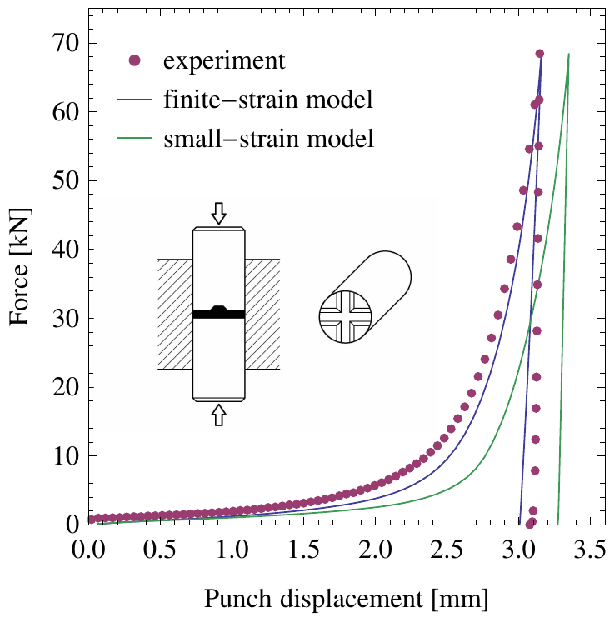} \\[1ex]
      ~~~~~~{\small (a)} & & ~~~~~~{\small (b)}
    \end{tabular}
    }
  \caption{Predictions of the small- and finite-strain versions of the PBG model
           for compression of alumina powder by:
           (a) flat punch (uniform uniaxial strain),
           (b) cross-shaped punch (nonuniform 3D deformation).
           The results are taken from \cite{StupPicBig14}.
           \label{fig:compare}}
\end{figure}

In the present paper, the model for cold forming of ceramic powders proposed by
Piccolroaz et al.~\cite{Piccolroaz06-1,Piccolroaz06-2} (called \lq PBG model' in the
following) is developed for large strain analyses, implemented in the finite element
method and numerically tested. The need for this large-strain generalization is
related to the fact that during ceramic forming the mean strain can easily reach
50\%, while peaks can touch 80\%. The differences between a small strain and a large
strain analysis can be appreciated from Fig.~\ref{fig:compare}, where small-strain
and large-strain predictions are reported for the force/displacement relation at the
top of a rigid mould containing an alumina ceramic powder. Results (taken from
\cite{StupPicBig14} and pertaining to a flat punch and to a punch with a \lq
cross-shaped' groove, respectively reported in Fig.~\ref{fig:compare}a and
\ref{fig:compare}b) clearly show that the large-strain analyses are more consistent
and in closer agreement with experimental results than the analyses performed under
the small strain hypothesis.

The model for powder compaction can be considered as paradigmatic of the
difficulties that can be encountered in the implementation of models for
geomaterials, since many \lq unconventional' features of plasticity are
simultaneously present to describe the complex transition from a loose granular
material (the powder) to a fully dense ceramic (the green body). These difficulties
enclose: (i.) the pressure-sensitive, $J_3$-dependent yield function introduced by
Bigoni and Piccolroaz~\cite{BigoniPiccolroaz04} (\lq BP yield function' in the
following), which is defined $+\infty$ in some regions outside the elastic domain;
(ii.) a nonlinear elastic behaviour even at small strain, (iii.) changes in elastic
response coupled to plastic deformation (elastoplastic coupling).

In this work, incremental (finite-step) constitutive equations are developed and
implemented for the finite-deformation version of the PBG model. In order to improve
the computational efficiency, the original model \cite{Piccolroaz06-2} is slightly
modified, but its essential features, including the elastoplastic coupling, are
preserved. Note that a consistent finite-element implementation of the elastoplastic
coupling at finite strain has not been reported in the literature so far. The model
is applied to simulate ceramic powder compaction with account for frictional contact
interaction.

The above-mentioned implementation difficulties are efficiently handled by using an
advanced hybrid symbolic-numeric approach implemented in \emph{AceGen}, a symbolic
code generation system \cite{Korelc02,Korelc09}. \emph{AceGen} combines symbolic and
algebraic capabilities of \emph{Mathematica}, automatic differentiation (AD)
technique, simultaneous optimization of expressions and automatic generation of
computer codes, and it is an efficient tool for rapid prototyping of numerical
procedures as well as for generation of highly optimized compiled codes (such as
finite element subroutines). Finite element computations have been carried out using
\emph{AceFEM}, a highly flexible finite element code that is closely integrated with
\emph{AceGen}.

Selected results of 2D and 3D simulations of powder compaction processes have
already been reported in \cite{StupPicBig14}, and the model predictions have been
compared to experimental data showing satisfactory agreement, see for instance
Fig.~\ref{fig:compare}. However, the finite-strain formulation and the numerical
strategies adopted for its implementation have not been presented in
\cite{StupPicBig14}, as that paper was aimed at providing an overview of
elastoplastic coupling in powder compaction processes. In the present paper, we
provide the details of the formulation and implementation, and, as an application,
we study the effect of friction and initial aspect ratio on compaction of alumina
powder in a cylindrical die.

\section{PBG model at small strain}


The small-strain PBG model \cite{Piccolroaz06-1} is briefly described below as a
reference for its finite-strain version introduced in the next section, with a
slight modification to the notation to make it more convenient for the subsequent
extension to the finite-strain framework. The model is fully defined by specifying
the free energy, the yield condition, and the plastic flow rule, and these are
provided below. For the details, including justification of the specific
constitutive assumptions and calibration of the model for alumina powder, refer to
Piccolroaz et al.~\cite{Piccolroaz06-1}.

\subsection{Free energy}

The total strain $\bfs{\varepsilon}$ is decomposed into the elastic
$\bfs{\varepsilon}_e$ and plastic $\bfs{\varepsilon}_p$ parts,
  \begin{equation}
    \bfs{\varepsilon} = \bfs{\varepsilon}_e + \bfs{\varepsilon}_p ,
  \end{equation}
and the free energy is assumed in the following form,
  \begin{equation}
    \label{eq:phi:ss}
    \phi(\bfs{\varepsilon},\bfs{\varepsilon}_p,p_c) =
      c \, \tr\bfs{\varepsilon}_e +
      (p_0+c) \left[ \left( d - \frac{1}{d} \right)
      \frac{(\tr\bfs{\varepsilon}_e)^2}{2\tilde{\kappa}} +
      d^{1/n} \tilde{\kappa} \exp \left(
        - \frac{\tr\bfs{\varepsilon}_e}{d^{1/n} \tilde{\kappa}} \right)
      \right] +
      \mu \tr\bfs{\varepsilon}_e^2
      -\frac{\mu}{3}(\tr\bfs{\varepsilon}_e)^2 ,
  \end{equation}
where the plastic strain $\bfs{\varepsilon}_p$ and the forming pressure $p_c$ are
adopted as internal variables, and
$\bfs{\varepsilon}_e=\bfs{\varepsilon}-\bfs{\varepsilon}_p$. The \emph{elastoplastic
coupling} is here introduced through the dependence of cohesion $c$, parameter $d$
and shear modulus $\mu$ on the forming pressure $p_c$, namely
  \begin{equation}
    \label{eq:hard:2}
    c = c_\infty [ 1 - \exp ( -\Gamma \langle p_c - p_{cb} \rangle ) ] ,
  \end{equation}
  \begin{equation}
    \label{eq:epc:1}
    d = 1 + B \langle p_c - p_{cb} \rangle ,
  \end{equation}
  \begin{equation}
    \label{eq:epc:2}
    \mu = \mu_0 + c \left( d - \frac{1}{d} \right) \mu_1 ,
  \end{equation}
where $\langle\cdot\rangle$ denotes the Macauley brackets operator,
$\tilde{\kappa}=\kappa/(1+e_0)$ and $\kappa$, $e_0$, $p_0$, $n$, $c_\infty$,
$\Gamma$, $p_{cb}$, $B$, $\mu_0$ and $\mu_1$ are material parameters. Note that the
elastoplastic coupling is related to the variation in $d$, so that, if $d$ remains
constant and equal to one, the elastic properties of the material remain unchanged
during plastic flow.


The forming pressure $p_c$ is assumed to depend on the volumetric part of the
plastic strain through the following relationship
  \begin{equation}
    \label{eq:hard:1}
    \tr\bfs{\varepsilon}_p = {\cal H}(p_c) , \qquad
    {\cal H}(p_c) =
      -\tilde{a}_1 \exp\left( -\frac{\Lambda_1}{p_c} \right)
      -\tilde{a}_2 \exp\left( -\frac{\Lambda_2}{p_c} \right) ,
  \end{equation}
where $\tilde{a}_i$ and $\Lambda_i$ are material parameters. In view of the above
dependence, the free energy could formally be expressed solely in terms of the total
strain $\bfs{\varepsilon}$ and the plastic strain $\bfs{\varepsilon}_p$. However,
the dependence of $p_c$ on $\bfs{\varepsilon}_p$ is implicit, i.e., $p_c$ cannot be
expressed as an explicit function of $\bfs{\varepsilon}_p$. It is thus convenient to
keep $p_c$ as an internal variable with the additional constraint introduced by
Eq.~(\ref{eq:hard:1}), see Section~\ref{sec:update}.


Note that the forming pressure $p_c$ and the cohesion $c$ are also used to define
the yield surface (see below). The corresponding governing equations
(\ref{eq:hard:1}) and (\ref{eq:hard:2}) are thus called `hardening laws' in
\cite{Piccolroaz06-1}.

\subsection{Inelastic strain rate}

The elastoplastic coupling, introduced above through the dependence of the free
energy $\phi$ on the plastic strain $\bfs{\varepsilon}_p$ and the forming pressure
$p_c$, is a crucial feature of the model. As a result, the stress depends not only
on the elastic strain, but also on the internal variables. Indeed, the stress
$\bfs{\sigma}$ is defined by
  \begin{equation}
    \bfs{\sigma} = \frac{\partial\phi}{\partial\bfs{\varepsilon}} ,
  \end{equation}
and its rate involves the contributions due to the evolution of the internal
variables,
  \begin{equation}
    \dot{\bfs{\sigma}} =
      \frac{\partial\bfs{\sigma}}{\partial\bfs{\varepsilon}} [\dot{\bfs{\varepsilon}}] +
      \frac{\partial\bfs{\sigma}}{\partial\bfs{\varepsilon}_p} [\dot{\bfs{\varepsilon}}_p] +
      \frac{\partial\bfs{\sigma}}{\partial p_c} \dot{p}_c =
      \mathbbm{E} [\dot{\bfs{\varepsilon}}] +
      \mathbbm{P} [\dot{\bfs{\varepsilon}}_p] +
      \bfm{P} \dot{p}_c =
      \mathbbm{E} [\dot{\bfs{\varepsilon}} - \dot{\bfs{\varepsilon}}_{\it in}] .
  \end{equation}
Here, $\mathbbm{E}$ is the elastic fourth-order tensor, $\mathbbm{P}$ and $\bfm{P}$
describe the elastoplastic coupling, and the inelastic strain rate
$\dot{\bfs{\varepsilon}}_{\it in}$ is defined as
  \begin{equation}
    \dot{\bfs{\varepsilon}}_{\it in} =
      - \mathbbm{E}^{-1} \mathbbm{P} [\dot{\bfs{\varepsilon}}_p]
      - \mathbbm{E}^{-1} [\bfm{P} \dot{p}_c] .
  \end{equation}
The inelastic strain rate $\dot{\bfs{\varepsilon}}_{\it in}$ is thus not equal to
the plastic strain rate $\dot{\bfs{\varepsilon}}_p$, and the former will be used in
the plastic flow rule, which is crucial for a consistent treatment of the
elastoplastic coupling, see Bigoni~\cite{Bigoni00}. The model is rate-independent,
hence by the time we understand here a time-like load parameter.

\subsection{Yield condition}

The yield condition is defined using the Bigoni--Piccolroaz (BP) yield function
\cite{BigoniPiccolroaz04}
  \begin{equation}
    \label{eq:F}
    F(\bfs{\sigma},p_c) = f(p,p_c) + \frac{q}{g(\theta)} \leq 0 ,
  \end{equation}
where
  \begin{equation}
    \label{eq:f:BP}
    f(p,p_c) = \left\{
      \begin{array}{ll}
        -M p_c \sqrt{(\Phi-\Phi^m)[2(1-\alpha)\Phi+\alpha]} \;\; & \mbox{if} \; \Phi \in [0,1], \\
        +\infty & \mbox{otherwise},
      \end{array} \right. \quad
    \Phi = \frac{p+c}{p_c+c} ,
  \end{equation}
  \begin{equation}
    \label{eq:g:BP}
    g(\theta) = \frac{1}{\cos[\beta\pi/6 - (1/3) \cos^{-1}(\gamma \cos 3\theta)]} ,
  \end{equation}
and $p$, $q$ and $\theta$ are the usual invariants of the stress tensor,
  \begin{equation}
    \label{eq:pqtheta}
    p = -\frac{1}{3} \tr\bfs{\sigma} , \qquad
    q = \sqrt{3 J_2} , \qquad
    \theta = \frac{1}{3} \cos^{-1} \left( \frac{3\sqrt{3}}{2} \,
      \frac{J_3}{J_2^{3/2}} \right) ,
  \end{equation}
  \begin{equation}
    \label{eq:J2J3}
    J_2 = \frac{1}{2} \tr(\dev\bfs{\sigma})^2 , \qquad
    J_3 = \frac{1}{3} \tr(\dev\bfs{\sigma})^3 , \qquad
    \dev\bfs{\sigma} = \bfs{\sigma} - \frac{1}{3} (\tr\bfs{\sigma}) \bfm{I} .
  \end{equation}
The forming pressure $p_c$ and the cohesion $c$, which depends on $p_c$ through
Eq.~(\ref{eq:hard:2}), define the size of the yield surface $F=0$ and its position
along the hydrostatic axis. Parameters $M$, $m$, $\alpha$, $\beta$ and $\gamma$
define the shape of the yield surface and are assumed constant.

It is seen from Eq.~(\ref{eq:f:BP}) that the BP yield function $F$ is defined
infinity for $p\not\in[-c,p_c]$, so it cannot be evaluated numerically for an
arbitrary stress state, and incremental schemes employing, for instance, the return
mapping algorithm cannot be applied directly. Therefore, following Stupkiewicz et
al.~\cite{StupDenPicBig14:sub}, an alternative implicit yield function $F^\ast$ is
used in practice, which has the same zero level set $F^\ast=0$ as the original yield
function (i.e., $F=0$) but behaves well for arbitrary stress states, see
Section~\ref{sec:implicit}.

\subsection{Plastic flow rule}


The flow rule is expressed in terms of the inelastic strain rate
$\dot{\bfs{\varepsilon}}_{\it in}$ rather than the plastic strain rate
$\dot{\bfs{\varepsilon}}_p$, see Bigoni~\cite{Bigoni00},
  \begin{equation}
    \label{eq:flow}
    \dot{\bfs{\varepsilon}}_{\it in} = \dot{\lambda} \hat{\bfm{n}} , \qquad
    \hat{\bfm{n}} = \bfm{n} - \frac{1}{3} \, \epsilon (1-\Phi)
      (\tr\bfm{n}) \bfm{I} , \qquad
    \bfm{n} = \frac{\partial F}{\partial\bfs{\sigma}} ,
  \end{equation}
where $\dot{\lambda}$ is the plastic multiplier satisfying the usual complementarity
conditions,
  \begin{equation}
    \label{eq:compl}
    \dot{\lambda} \geq 0 , \qquad
    F \leq 0 , \qquad
    \dot{\lambda} F = 0 .
  \end{equation}
Here, $\epsilon$ is a parameter that governs non-associativity of the flow rule
($0\leq\epsilon<1$), and $\epsilon=0$
corresponds to the associated flow rule.


\section{PBG model at finite strain}
\label{sec:finite}

The PBG model~\cite{Piccolroaz06-1} has been extended to the finite-strain framework
by the same group of authors in~\cite{Piccolroaz06-2}. In that model, the usual
multiplicative decomposition of the deformation gradient has been adopted, the free
energy has been expressed in terms of the logarithmic elastic strain while keeping
the same form (\ref{eq:phi:ss}) of the free energy function, and the BP yield
condition has been expressed in terms of the Biot stress tensor referred to the
initial configuration. With regard to the elastoplastic coupling and plastic flow
rule, the Biot stress and its conjugate strain measure have been used to define the
inelastic strain rate (using the general framework developed by
Bigoni~\cite{Bigoni00}), and that inelastic strain rate has subsequently been used
in the plastic flow rule. Finally, a complete set of rate equations has been
derived; however, incremental formulation and it finite element implementation have
not been attempted.


In this section, a finite-strain
formulation of the PBG model is introduced, which is more convenient for the
finite-element implementation than the formulation of Piccolroaz et
al.~\cite{Piccolroaz06-2}. At the same time, the essential features of that model
are preserved, namely the specific form of the free energy function, the
elastoplastic coupling framework of Bigoni~\cite{Bigoni00}, the BP yield
condition~\cite{BigoniPiccolroaz04}, and the plastic flow rule. The main difference
is in the selection of the internal variables and in the choice of the stress and
strain measures used to define the inelastic strain rate and to formulate the flow
rule. Also, the yield condition is here expressed in terms of the second
Piola-Kirchhoff stress referred to the intermediate configuration rather than in
terms of the Biot stress referred to the initial reference configuration which seems
more consistent with respect to the experimental testing procedures that are
typically used to calibrate the model.

\subsection{Free energy}

The deformation gradient $\bfm{F}$ is multiplicatively split into elastic
$\bfm{F}_e$ and plastic $\bfm{F}_p$ parts,
  \begin{equation}
    \bfm{F} = \bfm{F}_e \bfm{F}_p ,
  \end{equation}
and the following standard
kinematic quantities are introduced,
  \begin{equation}
    \label{eq:CCpbe}
    \bfm{C} = \bfm{F}^T \bfm{F} , \qquad
    \bfm{C}_p = \bfm{F}_p^T \bfm{F}_p , \qquad
    \bfm{b}_e = \bfm{F}_e \bfm{F}_e^T = \bfm{F} \bfm{C}_p^{-1} \bfm{F}^T ,
  \end{equation}
respectively, the total and plastic right Cauchy--Green tensors, and the elastic
left Cauchy--Green tensor. Furthermore, we have
  \begin{equation}
    J = J_e J_p , \quad
    J = \det\bfm{F} , \quad
    J_e = \det\bfm{F}_e = (\det\bfm{b}_e)^{1/2} , \quad
    J_p = \det\bfm{F}_p = (\det\bfm{C}_p)^{1/2} .
  \end{equation}
In order to conveniently treat the volumetric strains, which are essential in
modeling of powder compaction, the logarithmic elastic and plastic strain tensors
are introduced,
  \begin{equation}
    \label{eq:log}
    \bfs{\epsilon}_e = \log\bfm{V}_e = \frac{1}{2} \log\bfm{b}_e , \qquad
    \bfm{E}_p^{(0)} = \log\bfm{U}_p = \frac{1}{2} \log\bfm{C}_p ,
  \end{equation}
where $\bfm{F}_e=\bfm{V}_e\bfm{R}_e$, $\bfm{b}_e=\bfm{V}_e^2$,
$\bfm{F}_p=\bfm{R}_p\bfm{U}_p$, and $\bfm{C}_p=\bfm{U}_p^2$. The well-known benefit
of using the logarithmic strain measure is that the volumetric strain is simply
obtained as a trace of the corresponding strain tensor, and the total volumetric
strain is additively decomposed into elastic and plastic contributions.

Following Piccolroaz et al.~\cite{Piccolroaz06-2}, the free energy can be assumed in
the same functional form as in the small-strain model, Eq.~(\ref{eq:phi:ss}), with
the infinitesimal elastic strain $\bfs{\varepsilon}_e$ simply replaced by the
logarithmic strain $\bfs{\epsilon}_e$. However, this form is not efficient in
numerical implementation, and a modified free energy function is adopted in this
work. For completeness, application of the original free energy of Piccolroaz et
al.~\cite{Piccolroaz06-2} is discussed in Appendix~\ref{app:free}.

In the modified free energy function, the volumetric behavior is described in terms
of the logarithmic elastic strain $\bfs{\epsilon}_e$, just like in the original
model \cite{Piccolroaz06-2}, while the shear behavior is described by the term of
the neo-Hookean type formulated for the isochoric part of the elastic left
Cauchy--Green tensor $\bfm{b}_e$, namely
  \begin{equation}
    \label{eq:phi:neo}
    \phi(\bfm{C},\bfm{C}_p,p_c) =
      c \, \tr\bfs{\epsilon}_e +
      (p_0+c) \left[ \left( d - \frac{1}{d} \right)
      \frac{(\tr\bfs{\epsilon}_e)^2}{2\tilde{\kappa}} +
      d^{1/n} \tilde{\kappa} \exp \left(
        - \frac{\tr\bfs{\epsilon}_e}{d^{1/n} \tilde{\kappa}} \right)
      \right] +
      \frac{1}{2} \mu (\bar{I}_2 - 3) ,
  \end{equation}
where
  \begin{equation}
    \bar{I}_2 = \tr\bar{\bfm{b}}_e = J_e^{-2/3} \tr\bfm{b}_e , \qquad
    \det\bar{\bfm{b}}_e = 1 .
  \end{equation}
The right Cauchy--Green tensor $\bfm{C}$ is adopted as a relevant measure of the
total strain in view of the standard objectivity argument, and the plastic right
Cauchy--Green tensor $\bfm{C}_p$ is adopted as an internal variable. Since elastic
strains are here relatively small, the present modification of the free energy
function with respect to that of \cite{Piccolroaz06-2} does not noticeably affect
the actual elastic response.

The free energy (\ref{eq:phi:neo}) involves two invariants characterizing the
elastic strain, $\tr\bfs{\epsilon}_e$ and $\bar{I}_2$, that can be easily expressed
in terms of $\bfm{C}$ and $\bfm{C}_p$. Indeed, in view of (\ref{eq:CCpbe}) and
(\ref{eq:log}), we have
  \begin{equation}
    \label{eq:Je2}
    J_e^2 = ( \det \bfm{C} ) ( \det \bfm{C}_p^{-1} ) , \qquad
    \tr \bfm{b}_e = \tr ( \bfm{C} \bfm{C}_p^{-1} ) ,
  \end{equation}
so that
  \begin{equation}
    \label{eq:tree}
    \tr\bfs{\epsilon}_e = \frac{1}{2} \log J_e^2 , \qquad
    \bar{I}_2 = (J_e^2)^{-1/3} \tr ( \bfm{C} \bfm{C}_p^{-1} ) .
  \end{equation}

Parameters $c$, $d$ and $\mu$ in the free energy (\ref{eq:phi:neo}) are assumed to
depend on the forming pressure $p_c$ through
Eqs.~(\ref{eq:hard:2})--(\ref{eq:epc:2}), exactly as in the small-strain model,
while the forming pressure $p_c$ is related to $\bfm{C}_p$ by
  \begin{equation}
    \label{eq:hard:1:fs}
    (\det\bfm{C}_p)^{1/2} - 1 = {\cal H}(p_c) .
  \end{equation}
The above relationship is a consistent generalization of Eq.~(\ref{eq:hard:1})$_1$
to the finite deformation regime, where function ${\cal H}(p_c)$ is specified by
Eq.~(\ref{eq:hard:1})$_2$.

\subsection{Inelastic strain rate}

The inelastic strain rate and subsequently the flow rule are introduced using the
Green strain tensor $\bfm{E}^{(2)}=\frac{1}{2}(\bfm{C}-\bfm{I})$ and its conjugate
stress tensor, the second Piola--Kirchhoff stress $\bfm{T}^{(2)}$. This is a
particularly convenient choice because the second Piola--Kirchhoff stress
$\bfm{T}^{(2)}$ is directly obtained as the derivative of the free energy with
respect to $\bfm{C}$ using the following standard relationship:
  \begin{equation}
    \label{eq:T}
    \bfm{T}^{(2)} = \frac{\partial\phi}{\partial\bfm{E}^{(2)}}
      = 2 \frac{\partial\phi}{\partial\bfm{C}} .
  \end{equation}
Clearly, the material response is invariant with respect to the choice of
a pair of conjugate strain and stress measures, see \cite{HillRice73,Bigoni00}.
Evaluation of the rate of $\bfm{T}^{(2)}$ defines the inelastic strain rate
$\dot{\bfm{E}}_{\it in}$  according
to
  \begin{equation}
    \label{eq:Tdot}
    \dot{\bfm{T}}{}^{(2)} =
      \frac{\partial\bfm{T}^{(2)}}{\partial\bfm{E}^{(2)}} [\dot{\bfm{E}}{}^{(2)}] +
      \frac{\partial\bfm{T}^{(2)}}{\partial\bfm{C}_p} [\dot{\bfm{C}}_p] +
      \frac{\partial\bfm{T}^{(2)}}{\partial p_c} \dot{p}_c =
      \mathbbm{E} [\dot{\bfm{E}}{}^{(2)}] +
      \mathbbm{P} [\dot{\bfm{C}}_p] +
      \bfm{P} \dot{p}_c =
      \mathbbm{E} [\dot{\bfm{E}}{}^{(2)} - \dot{\bfm{E}}_{\it in}] ,
  \end{equation}
where
$\mathbbm{E}=\partial\bfm{T}^{(2)}/\partial\bfm{E}^{(2)}=2\partial\bfm{T}^{(2)}/\partial\bfm{C}$,
$\mathbbm{P}=\partial\bfm{T}^{(2)}/\partial\bfm{C}_p$,
$\bfm{P}=\partial\bfm{T}^{(2)}/\partial p_c$, and
  \begin{equation}
    \label{eq:Ein}
    \dot{\bfm{E}}_{\it in} =
      - \mathbbm{E}^{-1} \mathbbm{P} [\dot{\bfm{C}}_p]
      - \mathbbm{E}^{-1} [\bfm{P} \dot{p}_c] .
  \end{equation}


\subsection{Yield condition}

The yield condition is assumed to be defined by the BP yield function (\ref{eq:F})
expressed in terms of the second Piola-Kirchhoff stress $\bfm{T}^{(2)}_e$ referred
to the intermediate configuration,
  \begin{equation}
    \bfm{T}^{(2)}_e = J_e\bfm{F}_e^{-1} \bfs{\sigma} \bfm{F}_e^{-T}
       = J_p^{-1} \bfm{F}_p \bfm{T}^{(2)} \bfm{F}_p^T ,
  \end{equation}
where $\bfs{\sigma}=J^{-1}\bfm{F}\bfm{T}^{(2)}\bfm{F}^T$ is the Cauchy stress, so
that we have
  \begin{equation}
    F(\bfm{T}^{(2)}_e,p_c) \leq 0 ,
  \end{equation}
and the yield function $F$ is now defined by Eq.~(\ref{eq:F}) through the invariants
of $\bfm{T}^{(2)}_e$,
  \begin{equation}
    p = -\frac{1}{3} \tr\bfm{T}^{(2)}_e , \qquad
    J_2 = \frac{1}{2} \tr( \dev\bfm{T}^{(2)}_e )^2 , \qquad
    J_3 = \frac{1}{3} \tr( \dev\bfm{T}^{(2)}_e )^3 .
  \end{equation}
The above invariants, and thus the yield function, can be explicitly expressed in
terms of the second Piola-Kirchhoff stress $\bfm{T}^{(2)}$ and the plastic right
Cauchy--Green tensor $\bfm{C}_p$, the latter playing the role of a hardening
variable. Indeed, we have
  \begin{equation}
    \label{eq:pqtheta:fs}
    p = -\frac{1}{3} J_p^{-1} \tr(\bfm{T}^{(2)} \bfm{C}_p ) , \quad
    J_2 = \frac{1}{2} J_p^{-2} \tr[ \dev(\bfm{T}^{(2)} \bfm{C}_p ) ]^2 , \quad
    J_3 = \frac{1}{3} J_p^{-3} \tr[ \dev(\bfm{T}^{(2)} \bfm{C}_p ) ]^3 ,
  \end{equation}
which is easily verified in view of the following identity holding for
$n=1,2,\ldots$,
  \begin{equation}
    \tr(\bfm{T}^{(2)}_e)^n=J_p^{-n}\tr(\bfm{T}^{(2)}\bfm{C}_p)^n ,
  \end{equation}
and a similar identity holding for the respective deviators.

Note that the yield function was expressed in \cite{Piccolroaz06-2} in terms of the
Biot stress tensor referred to the initial configuration. The present choice of
$\bfm{T}^{(2)}_e$ is motivated by the typical model calibration procedures which are
based on the Cauchy stress or the nominal stress referred to the intermediate
configuration, see, for instance, \cite{MeschkeLiu99}. Considering that the elastic
strains are relatively small, the stress tensor $\bfm{T}^{(2)}_e$ is, in a sense
`close' to the Cauchy stress tensor $\bfs{\sigma}$, hence provides a physically
sound description of the yield surface. At the same time, as shown above, the yield
function depending on $\bfm{T}^{(2)}_e$ can be equivalently expressed solely in
terms of $\bfm{T}^{(2)}$ and $\bfm{C}_p$ which is not possible if the Cauchy stress
is used instead.\footnote
 {If the yield function is directly defined in terms of the Cauchy stress $\bfs{\sigma}$,
 it has to additionally depend on the total strain, for instance, on $\bfm{C}$, so that
 $$
   F(\bfs{\sigma}) = \tilde{F}(\bfm{T}^{(2)}, \bfm{C}_p, \bfm{C} ).
 $$
 If now Prager's consistency, $\dot{F}=0$,  is imposed, the term
 $$
   \frac{\partial \tilde{F}}{\partial \bfm{C}} \cdot \dot{\bfm{C}}
 $$
 yields an unsymmetrizing contribution to the tangent constitutive operator.
 Therefore, the choice of the Cauchy stress in the yield function leads to a model
 which does not fit the elastoplasticity framework of \cite{HillRice73,Bigoni00}.}

\subsection{Plastic flow rule}

The flow rule is expressed in terms of the inelastic strain rate $\dot{\bfm{E}}_{\it
in}$ defined in Eq.~(\ref{eq:Ein}) and the plastic flow direction tensor
$\hat{\bfm{N}}$ corresponding to the second Piola--Kirchhoff stress $\bfm{T}^{(2)}$.
The finite-strain counterpart of the flow rule (\ref{eq:flow}) is thus the
following,
  \begin{equation}
    \label{eq:flow:fs}
    \dot{\bfm{E}}_{\it in} = \dot{\lambda} \hat{\bfm{N}} , \qquad
    \hat{\bfm{N}} = \bfm{N} - \frac{1}{3} \, \epsilon (1-\Phi)
      \tr(\bfm{N}\bfm{C}_p^{-1}) \bfm{C}_p , \qquad
    \bfm{N} = \frac{\partial F}{\partial\bfm{T}^{(2)}} ,
  \end{equation}
with the usual complementarity conditions (\ref{eq:compl}). The term
$\tr(\bfm{N}\bfm{C}_p^{-1}) \bfm{C}_p$ in the formula for $\hat{\bfm{N}}$
corresponds to the volumetric part $(\tr\bfm{N}_e)\bfm{I}$ of the gradient
$\bfm{N}_e=\partial
F/\partial\bfm{T}^{(2)}_e=J_p\bfm{F}_p^{-T}\bfm{N}\bfm{F}_p^{-1}$ evaluated with
respect to $\bfm{T}^{(2)}_e$ in the intermediate configuration. Note that, as in the small-strain model, the associated
flow rule and thus the symmetry of the tangent operator are recovered for $\epsilon=0$.

\section{Finite element implementation}

The essential steps involved in derivation and implementation of incremental
constitutive relationships are presented in this section. The algorithmic treatment
employs the commonly-used backward-Euler integration scheme
and the classical return mapping algorithm, see, e.g.,
\cite{SimoHughes98,SouzaNeto08}. The present computer implementation has been
carried out using a symbolic code generation system \emph{AceGen} \cite{Korelc09},
and the related automation is also briefly discussed below.

\subsection{Implicit BP yield surface}
\label{sec:implicit}

The BP yield surface $F=0$ specified by Eqs.~(\ref{eq:F})--(\ref{eq:g:BP}) is highly
flexible considering the shape of its meridian and deviatoric sections. However,
this comes at the cost that the original yield function $F$ is not continuous and,
to enforce convexity, is defined infinity for $p\not\in[-c,p_c]$. As a result, the
BP yield function cannot be effectively evaluated for an arbitrary stress state so
that the classical return mapping algorithms cannot be directly applied. A general
strategy to overcome this problem, see Stupkiewicz et
al.~\cite{StupDenPicBig14:sub}, is to introduce an implicitly defined yield function
$F^\ast$ that has the same zero level set $F^\ast=0$ as the original yield function,
$F=0$, i.e., the same yield surface, but behaves well for arbitrary stress states.
The implicit yield function formulation is followed in this work and is briefly
described below. Alternative, less general approaches have been proposed in
\cite{BrannonLeelavanichkul10,Penasa14:sub}.

\begin{figure}[!htcb]
  \centerline{\inclps{0.4\textwidth}{!}{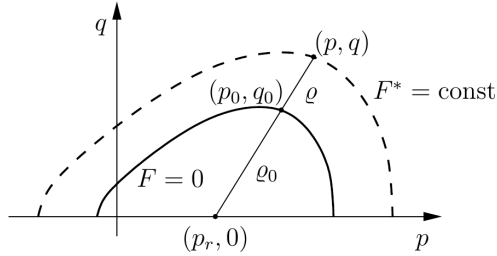}}
  \caption{Implicit yield function $F^\ast$ generated by a convex yield surface $F=0$.
           \label{fig:implicit}}
\end{figure}

Construction of a convex yield function $F^\ast(\bfs{\sigma},\cdot)$ generated by a
convex yield surface $F(\bfs{\sigma},\cdot)=0$
is illustrated in Fig.~\ref{fig:implicit}. Consider the $(p,q)$-space corresponding
to a fixed Lode angle $\theta$, and introduce a reference point $(p_r,0)$ inside the
yield surface $F=0$. Further, denote by $\varrho$ the distance between the reference
point $(p_r,0)$ and the current stress point $(p,q)$ and by $\varrho_0$ the distance
between the reference point $(p_r,0)$ and the image point $(p_0,q_0)$ that lies on
the yield surface $F=0$,
  \begin{equation}
    \varrho = \| \bfs{\varrho} \| , \qquad
    \varrho_0 = \| \bfs{\varrho}_0 \| , \qquad
    \bfs{\varrho} = (p-p_r,q) , \qquad
    \bfs{\varrho}_0 = (p_0-p_r,q_0) ,
  \end{equation}
and we have $\bfs{\varrho}_0/\varrho_0=\bfs{\varrho}/\varrho$. The yield function
$F^\ast$ is then defined by
  \begin{equation}
    \label{eq:Fs}
    F^\ast(\bfs{\sigma},\cdot) = \frac{\varrho}{\varrho_0} - 1 .
  \end{equation}
By construction, the yield function $F^\ast$ is convex and generates a family of
self-similar iso-surfaces $F^\ast=\const$.

In order to evaluate the yield function $F^\ast(\bfs{\sigma},\cdot)$ for an
arbitrary stress $\bfs{\sigma}$, a nonlinear equation must be solved to determine
$\varrho_0$. That equation corresponds to the condition that the image point
$(p_0,q_0)$ lies on the yield surface. The yield function $F^\ast$ is thus an
\emph{implicit} function. Consequently, its derivatives, for instance, the gradient
used in the flow rule, involve the derivatives of the implicit dependence of
$\varrho_0$ on the stress $\bfs{\sigma}$ and, possibly, also on hardening variables.
The details can be found in \cite{StupDenPicBig14:sub}.



The present implementation of the PBG model employs the above implicit formulation
of the BP yield function. Accordingly, the actual use of the implicit yield function
$F^\ast$ and its gradient $\bfm{N}^\ast$ is denoted below by a `$\ast$' in the
superscript.


\subsection{Incremental flow rule}

With regard to finite element implementation, the constitutive equations specified
in Section~\ref{sec:finite} must be cast in an incremental form, i.e., an
appropriate time integration scheme must be applied to the evolution equations for
internal variables.

Using the flow rule (\ref{eq:flow:fs}), the rate constitutive equation
(\ref{eq:Tdot}) is rewritten as
  \begin{equation}
    \label{eq:Tdot:flow}
    \dot{\bfm{T}}{}^{(2)} = \mathbbm{E} [\dot{\bfm{E}}{}^{(2)}
      - \dot{\lambda} \hat{\bfm{N}}{}^\ast ] ,
  \end{equation}
which upon application of the implicit backward-Euler integration scheme yields
  \begin{equation}
    \label{eq:deltaT}
    \bfm{T}^{(2)}_{n+1} - \bfm{T}^{(2)}_n = \mathbbm{E}_{n+1} [
      \bfm{E}^{(2)}_{n+1} - \bfm{E}^{(2)}_n - \Delta\lambda \hat{\bfm{N}}{}^\ast_{n+1} ] ,
  \end{equation}
where $n+1$ and $n$ in the subscript denote, respectively, the current time
$t=t_{n+1}$ and the previous time $t=t_n$, at which the corresponding quantities are
evaluated. The incremental
flow rule (\ref{eq:deltaT}) is accompanied by the complementarity conditions,
  \begin{equation}
    \label{eq:compl:inc}
    F^\ast_{n+1} = F^\ast(\bfm{T}^{(2)}_{e,n+1},p_{c,n+1}) \leq 0 , \qquad
    \Delta\lambda \geq 0 , \qquad
    \Delta\lambda F^\ast_{n+1} = 0 ,
  \end{equation}
that are enforced at the end of the time increment, consistently with the
backward-Euler scheme applied to integrate the rate equation (\ref{eq:Tdot:flow}).

Considering that arbitrary stress states can be encountered during iterative
solution of the return mapping algorithm, the plastic flow direction
$\hat{\bfm{N}}{}^\ast$ is modified according to
  \begin{equation}
    \label{eq:Nstar}
    \hat{\bfm{N}}{}^\ast = \bfm{N}^\ast - \frac{1}{3} \, \epsilon (1-\Phi_0)
      \tr(\bfm{N}^\ast\bfm{C}_p^{-1}) \bfm{C}_p , \qquad
    \Phi_0 = \frac{p_0+c}{p_c+c} , \qquad
    p_0 = \frac{p + F^\ast p_r}{1+F^\ast} ,
  \end{equation}
so that $\Phi_0\in[0,1]$, just like $\Phi\in[0,1]$ for the stresses satisfying
$F^\ast=0$. Of course, the flow rule is unaltered for $F^\ast=0$.

\paragraph{Remark 1.}
The simple backward-Euler scheme is usually avoided in finite-strain plasticity, in
particular, in the case of plastically incompressible (or nearly incompressible)
materials, e.g., in metal plasticity, and integration schemes employing the
exponential map are then preferable, cf.~\cite{WeberAnand90,SteinmannStein96}.
However, the present plastic flow rule, formulated in terms of the inelastic strain
rate $\dot{\bfm{E}}_{\it in}$ to consistently introduce the elastoplastic coupling,
is not well suited for application of an exponential map integrator.

\subsection{Constitutive update problem}
\label{sec:update}


In the constitutive update problem, given are the deformation gradient
$\bfm{F}_{n+1}$ at the current time step and the internal variables $\bfm{C}_{p,n}$
and $p_{c,n}$ at the previous time step, and the goal is to find the current
internal variables $\bfm{C}_{p,n+1}$ and $p_{c,n+1}$ and the plastic multiplier
$\Delta\lambda$ that satisfy the incremental flow rule (\ref{eq:deltaT}), the
complementarity conditions (\ref{eq:compl:inc}) and the constitutive relationship
between $p_{c,n+1}$ and $\bfm{C}_{p,n+1}$ specified by Eq.~(\ref{eq:hard:1:fs}).

The second Piola--Kirchhoff stress $\bfm{T}^{(2)}_{n+1}$, which is needed to
evaluate the yield function $F^\ast_{n+1}$ and its gradient $\bfm{N}^\ast_{n+1}$ in
Eqs.~(\ref{eq:deltaT})--(\ref{eq:Nstar}), is defined by the free energy according to
  \begin{equation}
    \bfm{T}^{(2)}_{n+1} = 2 \frac{\partial\phi(\bfm{C}_{n+1},\bfm{C}_{p,n+1},p_{c,n+1})}
      {\partial\bfm{C}_{n+1}} .
  \end{equation}

The constitutive update problem is solved here using the classical return-mapping
algorithm \cite{SimoHughes98,SouzaNeto08}. The trial stress is first computed by
assuming that the response is elastic,
  \begin{equation}
    \label{eq:trial:s}
    \bfm{T}_{n+1}^{(2)\it trial} =
      2 \frac{\partial\phi(\bfm{C}_{n+1},\bfm{C}_{p,n},p_{c,n})}
      {\partial\bfm{C}_{n+1}} ,
  \end{equation}
for which the trial value of the yield function is evaluated according to
  \begin{equation}
    \label{eq:trial:f}
    F^{\ast\it trial}_{n+1} = F^\ast (\bfm{T}_{e,n+1}^{(2)\it trial}, p_{c,n}) ,
  \end{equation}
where the necessary invariants of $\bfm{T}_{e,n+1}^{(2)\it trial}$ are expressed in
terms of $\bfm{T}_{n+1}^{(2)\it trial}$ and $\bfm{C}_{p,n}$ using the formulae in
Eq.~(\ref{eq:pqtheta:fs}).

If $F^{\ast\it trial}_{n+1}\leq0$ then the step is elastic, and
  \begin{equation}
    \bfm{C}_{p,n+1} = \bfm{C}_{p,n} , \qquad
    p_{c,n+1} = p_{c,n} , \qquad
    \Delta\lambda = 0 .
  \end{equation}

If $F^{\ast\it trial}_{n+1}>0$ then the step is plastic and a set of nonlinear
equations must be solved,
  \begin{equation}
    \label{eq:Qh0}
    \fem{Q}_{n+1}(\fem{h}_{n+1}) = \fem{0} ,
  \end{equation}
where the vector of unknowns $\fem{h}_{n+1}$ comprises the internal variables and
the plastic multiplier $\lambda$,
  \begin{equation}
    \fem{h}_{n+1} = \{ C_{p,11}, C_{p,22}, C_{p,33},
      C_{p,12}, C_{p,13}, C_{p,23}, \lambda, p_c \}_{n+1} ,
  \end{equation}
and $C_{p,ij}$ denote the components of $\bfm{C}_p$. The local residual vector
$\fem{Q}_{n+1}$ is defined as
  \begin{equation}
    \label{eq:Q}
    \fem{Q}_{n+1} = \{ {\cal Z}_{11}, {\cal Z}_{22}, {\cal Z}_{33}, {\cal Z}_{12},
      {\cal Z}_{13}, {\cal Z}_{23}, F^\ast_{n+1}, {\cal A}_{n+1} \} ,
  \end{equation}
where ${\cal Z}_{ij}$ are the component-wise residuals corresponding to the
incremental flow rule (\ref{eq:deltaT}),
  \begin{equation}
    {\cal Z}_{ij} = \left(
      \bfm{T}^{(2)}_{n+1} - \bfm{T}^{(2)}_n - \mathbbm{E}_{n+1} [
      \bfm{E}^{(2)}_{n+1} - \bfm{E}^{(2)}_n - \Delta\lambda \hat{\bfm{N}}{}^\ast_{n+1} ]
      \right)_{ij} ,
  \end{equation}
and ${\cal A}_{n+1}=0$ is the equation that relates the hardening variable
$p_{c,n+1}$ and the plastic strain $\bfm{C}_{p,n+1}$, cf.\ Eq.~(\ref{eq:hard:1:fs}),
  \begin{equation}
    \label{eq:A}
    {\cal A}_{n+1} = (\det\bfm{C}_{p,n+1})^{1/2} - 1 - {\cal H}(p_{c,n+1}) .
  \end{equation}
Equation (\ref{eq:Qh0}) is solved using the Newton method according to the following
iterative scheme:
  \begin{equation}
    \fem{h}_{n+1}^{(j+1)} = \fem{h}_{n+1}^{(j)} + \Delta\fem{h}_{n+1}^{(j)} , \qquad
    \Delta\fem{h}_{n+1}^{(j)} =
      -\left( \frac{\partial\fem{Q}_{n+1}}{\partial\fem{h}_{n+1}} \right)^{-1}
      \fem{Q}_{n+1}(\fem{h}_{n+1}^{(j)}) .
  \end{equation}

Once the constitutive update problem is solved, the first Piola--Kirchhoff stress
$\bfm{P}_{n+1}$ and the consistent constitutive tangent $\mathbbm{C}^{\it ep}_{n+1}$
are computed,
  \begin{equation}
    \bfm{P}_{n+1} = \bfm{F}_{n+1} \bfm{T}^{(2)}_{n+1} , \qquad
    \mathbbm{C}^{\it ep}_{n+1} = \frac{\partial\bfm{P}_{n+1}}{\partial\bfm{F}_{n+1}} ,
  \end{equation}
that are needed at the global level where the equilibrium equations are solved using
the finite element method. It is reminded here that the internal variables
$\fem{h}_{n+1}$ implicitly depend on $\bfm{F}_{n+1}$ through Eq.~(\ref{eq:Qh0}), and
the derivative of this implicit dependence must be accounted for when computing the
consistent tangent $\mathbbm{C}^{\it ep}_{n+1}$, cf.\ \cite{Michaleris94}.

In the present implementation, the standard return mapping algorithm specified above
has been actually replaced by a more robust algorithm employing the augmented primal
closest-point projection method proposed by Perez-Foguet and
Armero~\cite{PerezFoguetArmero02}. This improved algorithm is described in
Appendix~\ref{app:AL}.

The model has been implemented using \emph{AceGen} \cite{Korelc02,Korelc09}, a
symbolic code generation system that combines the symbolic capabilities of
\emph{Mathematica} (www.wolfram.com) with the automatic differentiation technique
and additional tools for optimization and automatic generation of computer codes.
The present formulation of incremental elastoplasticity and the structure of the
constitutive update problem fit the general formulation introduced by
Korelc~\cite{Korelc09}, hence the automation approach developed in~\cite{Korelc09}
can be directly applied to derive the necessary finite element routines. In
particular, the incremental constitutive model is fully defined by specifying the
local residual $\fem{Q}_{n+1}$ in terms of the internal variables $\fem{h}_{n+1}$,
as done above, while the remaining part of the formulation remains unaltered. The
details
are omitted here; an interested reader is referred to~\cite{Korelc09}, see also
Section~2 in \cite{KorelcStupkiewicz14} for a concise presentation of the present
automation approach in finite-strain elastoplasticity.

Application of the automatic differentiation technique implemented in \emph{AceGen}
results in the exact linearization of the incremental constitutive relationships
which is highly beneficial for the overall performance of the Newton-based
computational scheme applied to solve the global equilibrium equations.

It has been checked numerically that the consistent elastoplastic tangent
$\mathbbm{C}^{\it ep}_{n+1}$ corresponding to the associated flow rule, i.e., for
$\epsilon=0$, is not symmetric for a finite strain increment. However, the symmetry
is recovered for the strain increment decreasing to zero. This shows consistency of
the present incremental scheme with the rate formulation in which a symmetric
elastoplastic tangent is obtained (for associative plasticity) from the Prager's
consistency condition, see \cite{Piccolroaz06-2}.




\section{Numerical example}

An application of the model to finite element computations is presented in this
section. Specifically, compaction of an alumina powder in a cylindrical die is
considered with account for die-wall friction, and the effect of friction
coefficient and initial aspect ratio is studied in detail. Other examples, including
comparison to experimental results, can be found in \cite{StupPicBig14}.

Consider thus cold compression of alumina powder into a rigid cylindrical mould of
radius $r$. The powder specimen has an initial height $h_0$ and is compressed by a
rigid punch with a maximum force corresponding to the average compaction pressure of
160 MPa. Three values of the initial aspect ratio of the alumina sample are
employed, $h_0/r=2$, $4$ and $6$. Coulomb friction is assumed at the powder-mould
and powder-punch contact interfaces, and three values of the friction coefficient
are used, $\mu=0.1$, $0.3$ and $0.5$. Material parameters corresponding to alumina
powder are provided in Table~\ref{tab:parameters}, see
\cite{Piccolroaz06-1,StupPicBig14}. The value of parameter $\mu_0$ has been
increased with respect to that adopted in \cite{Piccolroaz06-1,StupPicBig14} since
it has been noticed that the latter may lead to exceedingly large elastic shear
strains. At the same time, in compression-dominated processes, such as those
considered in \cite{Piccolroaz06-1,StupPicBig14} and in the present paper, the
overall response is not significantly affected.

\begin{table}[!htcb]
  \caption{Material parameters for alumina powder.
           \label{tab:parameters}}
  \vspace{1ex}
  \centerline{\footnotesize
  \renewcommand{\arraystretch}{1.25}
\begin{tabular}{llllllllll} \hline
$M$ & $m$ & $\alpha$ & $\beta$ & $\gamma$ ~~~ & $\epsilon$ & $\tilde{a}_1$ & $\tilde{a}_2$ & $\Lambda_1$ (MPa)\!\! & $\Lambda_1$ (MPa)\!\! \\ \hline
1.1 & 2 & 0.1 & 0.19 & 0.9 & 0.5 & 0.383 & 0.124 & 1.8 & 40 \\ \hline
$c_\infty$ (MPa)\!\! & $\Gamma$ (MPa$^{-1}$)\!\! & $p_{cb}$ (MPa)\!\! & $B$ (MPa$^{-1}$)\!\! & $n$ & $\mu_0$ (MPa)\!\! & $\mu_1$ & $\kappa$ & $e_0$ & $p_0$ (MPa)\!\! \\ \hline
2.3 & 0.026 & 3.2 & 0.18 & 6 & 20 & 64 & 0.04 & 2.129 & 0.063 \\ \hline
\end{tabular}
  }
  \vspace{2ex}
\end{table}

In the finite element implementation, an axisymmetric under-integrated four-node
element employing the volumetric-deviatoric split and Taylor expansion of shape
functions \cite{KorelcWriggers97} is used for the solid, and the augmented
Lagrangian method used to enforce the frictional contact constraints
\cite{AlartCurnier91,LengKorStu11}.

\begin{figure}[!htcb]
  \centerline{
    \begin{tabular}{ccc}
      \inclps{0.32\textwidth}{!}{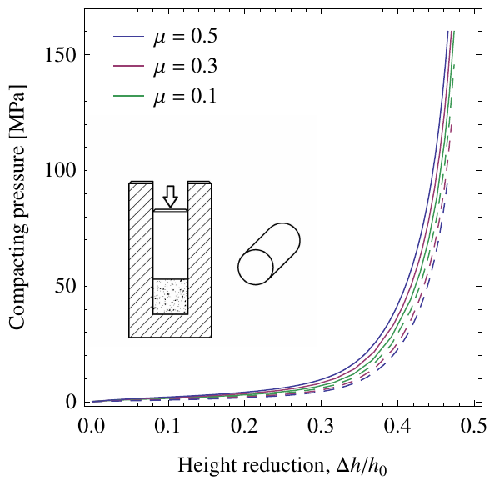} &
      \inclps{0.32\textwidth}{!}{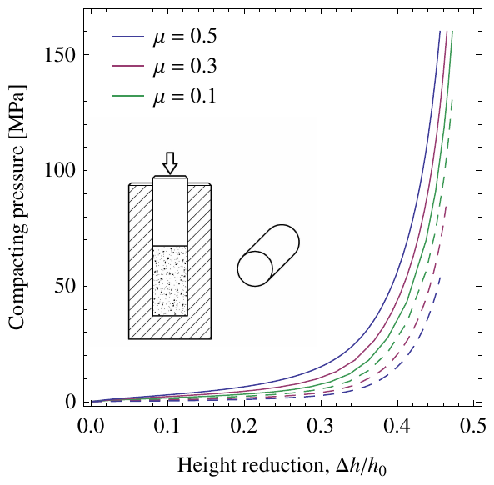} &
      \inclps{0.32\textwidth}{!}{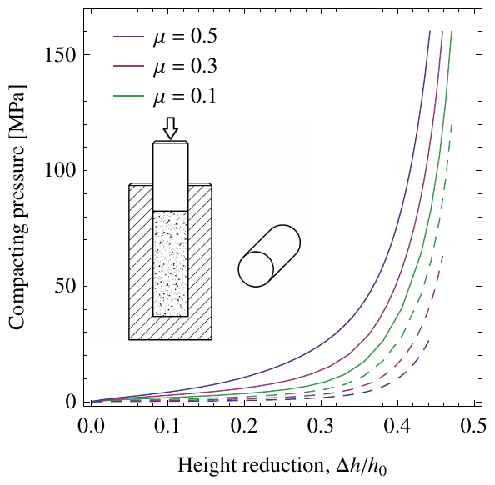} \\[1ex]
      ~~~~{\small (a)} & ~~~~{\small (b)} & ~~~~{\small (c)}
    \end{tabular}
    }
  \caption{The average compacting pressure (solid lines) and the average pressure
           at the bottom part of the mould (dashed lines) as a function of the height
           reduction $\Delta h/h_0$ for:
           (a) $h_0/r=2$, (b) $h_0/r=4$, (c) $h_0/r=6$.
    \label{fig:ax:pressure}}
\end{figure}

The effect of friction coefficient $\mu$ and initial aspect ratio $h_0/r$ is
illustrated in Fig.~\ref{fig:ax:pressure} which shows the average compacting
pressure as a function of the height reduction $\Delta h/h_0$. In each case, in
addition to the compacting pressure indicated by a solid line, the corresponding
average pressure at the bottom part of the mould is also shown using a dashed line
of the same color. Clearly, for frictionless contact at the die wall, the two
pressures would be equal one to the other, and the difference increases with
increasing friction coefficient $\mu$ and with increasing initial aspect ratio
$h_0/r$.

\begin{figure}[!htcb]
  \centerline{\inclps{0.65\textwidth}{!}{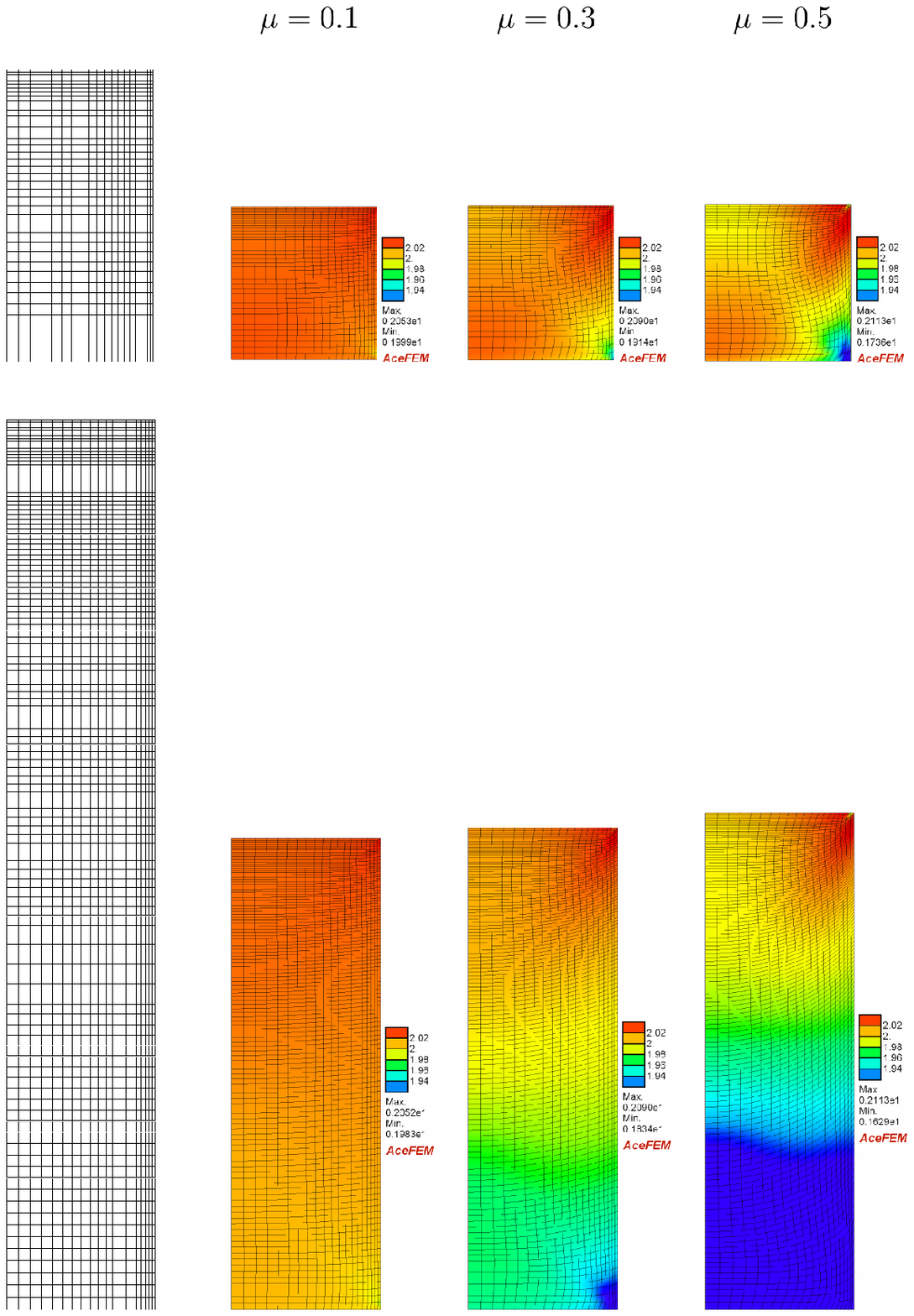}}
  \caption{Axisymmetric compression into a rigid cylindrical mould:
           the undeformed mesh (left) and the deformed mesh corresponding to
           a prescribed maximum force and different friction coefficients $\mu$
           for $h_0/r=2$ (upper row) and $h_0/r=6$ (lower row). The color map,
           identical for all figures, indicates the density $\varrho$ (in g/cm$^3$).
           \label{fig:ax:rho}}
\end{figure}

The finite element mesh representing one half of the cross-section of the
(axisymmetric) sample is shown in Fig.~\ref{fig:ax:rho} for $h_0/r=2$ and $h_0/r=6$.
The undeformed mesh is shown in the left column, and the deformed meshes
corresponding to the maximum compression force and different friction coefficients
are shown aside. The color map, identical for all figures, indicates the resulting
density of alumina powder. Due to friction, the density is nonuniform within the
cross-section of the specimen, and the substantial effect of friction coefficient
$\mu$ and initial aspect ratio $h_0/r$ on the density distribution is clearly seen
in Fig.~\ref{fig:ax:rho}.

The effect of friction, which is more pronounced for higher aspect ratios, results
in reduced overall compaction so that the final height corresponding to the same
prescribed maximum compression force depends on the friction coefficient. This is
particularly visible for $h_0/r=6$. Also, the deformation pattern is affected by the
shear stresses due to friction at the die wall, which is seen in the distortion of
the initially rectangular mesh.

\newpage

\begin{figure}[!htcb]
  \centerline{\inclps{0.65\textwidth}{!}{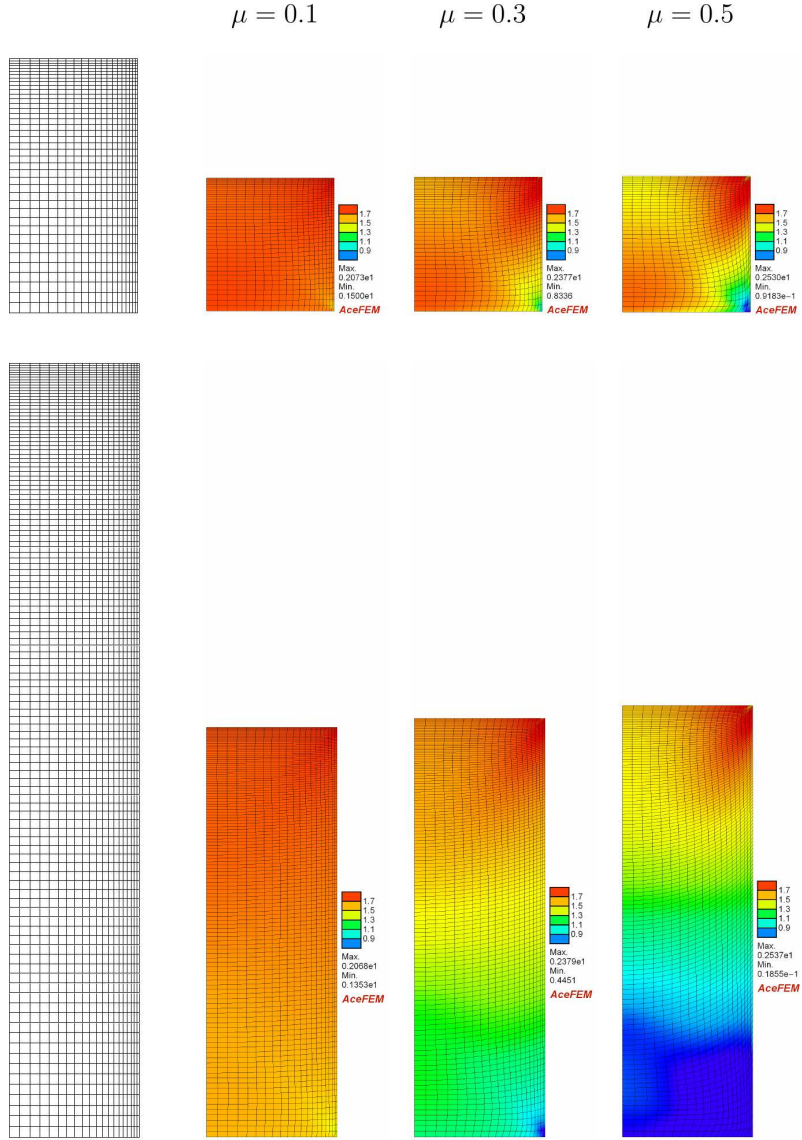}}
  \caption{Axisymmetric compression into a rigid cylindrical mould:
           the undeformed mesh (left) and the deformed mesh corresponding to
           a prescribed maximum force and different friction coefficients $\mu$
           for $h_0/r=2$ (upper row) and $h_0/r=6$ (lower row). The color map,
           identical for all figures, indicates the cohesion $c$ (in MPa).
           \label{fig:ax:c}}
\end{figure}

In the PBG model, plastic hardening is governed by the volumetric plastic
deformation. Nonuniform distribution of density results thus in nonuniform hardening
within the sample. This is illustrated in Fig.~\ref{fig:ax:c} which shows the
distribution of the cohesion $c$, again indicated by the same color map for all
figures. The pattern of inhomogeneity of $c$ is qualitatively similar to that shown
in Fig.~\ref{fig:ax:rho}, and the same applies to the distribution of the forming
pressure $p_c$ (not shown for brevity).


\section{Conclusion}

A finite strain model for powder compaction has been developed through the
derivation of an incremental scheme allowing the successful FE implementation of a
series of \lq non-standard' constitutive features, including: (i.) nonlinear elastic
behaviour even at small strain; (ii.) coupling between elastic and plastic
deformation; (iii.) pressure- and $J_3$-dependent yielding; (iv.) non-isochoric
flow. The extension of concepts established at small strain to the large deformation
context has required a proper selection of the stress variable to be employed for
the yield function and the definition of a neo-Hookean \lq correction' to the small
strain elastic potential. An incremental scheme compatible with the finite-strain
formulation of elastoplastic coupling has also been developed. The numerical tests
performed on the model have indicated a correct and robust behaviour of the
developed code and have demonstrated the possibility of accurate simulation of the
transition from a granular material to a fully dense body occurring during the cold
forming of ceramic powders.

\section*{Acknowledgments}
The authors gratefully acknowledge financial support from European Union FP7 project
under contract number PIAP-GA-2011-286110.

\appendix

\section{Free energy expressed in terms of the logarithmic elastic strain $\bfs{\epsilon}_e$}
\label{app:free}

In this appendix, we provide the formulation corresponding to the original form of
the free energy function as proposed in \cite{Piccolroaz06-2}. Consider thus the
small-strain free energy function (\ref{eq:phi:ss}) with the infinitesimal elastic
strain $\bfs{\varepsilon}_e$ simply replaced by the logarithmic elastic strain
$\bfs{\epsilon}_e$, viz.
  \begin{equation}
    \label{eq:phi:fs}
    \phi(\bfm{C},\bfm{C}_p,p_c) =
      c \, \tr\bfs{\epsilon}_e +
      (p_0+c) \left[ \left( d - \frac{1}{d} \right)
      \frac{(\tr\bfs{\epsilon}_e)^2}{2\tilde{\kappa}} +
      d^{1/n} \tilde{\kappa} \exp \left(
        - \frac{\tr\bfs{\epsilon}_e}{d^{1/n} \tilde{\kappa}} \right)
      \right] +
      \mu \tr\bfs{\epsilon}_e^2
      -\frac{\mu}{3}(\tr\bfs{\epsilon}_e)^2 .
  \end{equation}
The difference with respect to the free energy function (\ref{eq:phi:neo}) concerns
the shear response introduced by the last two terms in Eq.~(\ref{eq:phi:fs}). We
also note that, in the original form \cite{Piccolroaz06-2}, the logarithmic plastic
strain $\bfm{E}_p^{(0)}$ has been adopted as the internal variable rather than the
plastic right Cauchy--Green tensor $\bfm{C}_p$. The two tensors are related by
Eq.~(\ref{eq:log})$_2$, or by the inverse relationship
$\bfm{C}_p=\exp(2\bfm{E}_p^{(0)})$, hence both formulations are equivalent. As shown
below, explicit formulae involving $\bfm{C}_p$ are readily available, hence using
$\bfm{E}_p^{(0)}$ as the internal variable would introduce an additional and
unnecessary complexity related to the tensor exponential function relating
$\bfm{E}_p^{(0)}$ and $\bfm{C}_p$.


The free energy~(\ref{eq:phi:fs}) is expressed in terms of two invariants of the
logarithmic elastic strain $\bfs{\epsilon}_e$, namely $\tr\bfs{\epsilon}_e$ and
$\tr\bfs{\epsilon}_e^2$. Evaluation of the former does not pose any difficulties,
see Eqs.~(\ref{eq:Je2})--(\ref{eq:tree}). In order to compute the second invariant,
$\tr\bfs{\epsilon}_e^2$, the logarithmic elastic strain $\bfs{\epsilon}_e$ must be
computed explicitly (this has been avoided in the case of the first invariant
$\tr\bfs{\epsilon}_e$). In view of Eqs.~(\ref{eq:CCpbe})$_3$ and (\ref{eq:log})$_1$,
we have
  \begin{equation}
    \bfs{\epsilon}_e = \frac{1}{2} \log ( \bfm{F} \bfm{C}_p^{-1} \bfm{F}^T ) .
  \end{equation}
However, this expression involves the deformation gradient $\bfm{F}$ and not the
right Cauchy--Green tensor $\bfm{C}$. Note that an explicit dependence on $\bfm{C}$
(or equivalently on $\bfm{E}^{(2)}$) is needed in the present elastoplasticity
framework, see Eqs.~(\ref{eq:T})--(\ref{eq:Tdot}). Due to objectivity, the free
energy function is invariant to a rigid-body rotation, hence we have
  \begin{equation}
    \label{eq:tree2}
    \tr\bfs{\epsilon}_e^2 = \tr(\bfs{\epsilon}_e^\ast)^2 , \qquad
    \bfs{\epsilon}_e^\ast = \frac{1}{2} \log ( \bfm{U} \bfm{C}_p^{-1} \bfm{U} )
      = \frac{1}{2} \log ( \bfm{C}^{1/2} \bfm{C}_p^{-1} \bfm{C}^{1/2} ) ,
  \end{equation}
where $\bfs{\epsilon}_e^\ast$ refers to a special configuration rotated by
$\bfm{R}=\bfm{F}\bfm{U}^{-1}$ with respect to the current configuration. It is seen
that the resulting formula (\ref{eq:tree2}) for $\tr\bfs{\epsilon}_e^2$ is rather
complex as it involves the tensor logarithm function and the square root of
$\bfm{C}$. Even more importantly, if the above formulation was adopted, then
solution of the constitutive update problem would involve the third derivative the
tensor logarithm function, which would be associated with a prohibitively high
computational cost.

Considering that the elastic strains are relatively small in the materials of
interest, the second invariant $\tr\bfs{\epsilon}_e^2$ can be approximated with a
high accuracy by exploiting the following approximation of the logarithmic strain,
see~\cite{Bazant98},
  \begin{equation}
    \bfs{\epsilon}_e = \log \bfm{V}_e \approx \frac{1}{2} ( \bfm{V}_e - \bfm{V}_e^{-1} ) ,
  \end{equation}
which leads to
  \begin{equation}
   \label{eq:tree2apr}
   \tr\bfs{\epsilon}_e^2 \approx
      \frac{1}{4} ( \tr\bfm{b}_e + \tr\bfm{b}_e^{-1} - 6) =
      \frac{1}{4} [ \tr(\bfm{C}\bfm{C}_p^{-1}) + \tr(\bfm{C}\bfm{C}_p^{-1})^{-1} - 6] .
  \end{equation}

Concluding, the free energy function (\ref{eq:phi:fs}) can be directly expressed in
terms of $\bfm{C}$ and $\bfm{C}_p$ using Eq.~(\ref{eq:tree}) for
$\tr\bfs{\epsilon}_e$ and Eq.~(\ref{eq:tree2}) or (\ref{eq:tree2apr}) for
$\tr\bfs{\epsilon}_e^2$.

\section{Augmented primal closest-point projection}
\label{app:AL}

The augmented primal closest-point projection method proposed in
\cite{PerezFoguetArmero02} proved to significantly improve the convergence of the
Newton method used for the solution of the return-mapping equations in the
constitutive update problem defined in Section~\ref{sec:update}. The idea is to
apply the augmented Lagrangian method to enforce the inequality constraints
corresponding to the incremental complementarity conditions, cf.\
Eq.~(\ref{eq:compl:inc}),
  \begin{equation}
    F^\ast_{n+1} \leq 0 , \qquad
    \Delta\lambda \geq 0 , \qquad
    \Delta\lambda F^\ast_{n+1} = 0 .
  \end{equation}
The simple treatment proposed in \cite{PerezFoguetArmero02} amounts to replacing the
plastic multiplier $\Delta\lambda$ in the incremental flow rule (\ref{eq:deltaT}) by
the augmented one, $\Delta\hat{\lambda}$,
  \begin{equation}
    \label{eq:gammahat}
    \Delta\hat{\lambda} = \max (0, \Delta\lambda + \varrho F^\ast_{n+1}) ,
  \end{equation}
where $\varrho>0$. The condition $F^\ast_{n+1}=0$ is modified accordingly,
  \begin{equation}
    \hat{F}^\ast_{n+1} = \frac{1}{\varrho} (\Delta\hat{\lambda} - \Delta\lambda) = 0 ,
  \end{equation}
which now enforces $F^\ast_{n+1}=0$ when $\Delta\lambda+\varrho F^\ast_{n+1}>0$ and
$\Delta\lambda=0$ otherwise. As a result, the local residual $\fem{Q}_{n+1}$ is
redefined as
  \begin{equation}
    \fem{Q}_{n+1} = \{ {\cal Z}_{11}, {\cal Z}_{22}, {\cal Z}_{33}, {\cal Z}_{12},
      {\cal Z}_{13}, {\cal Z}_{23}, \hat{F}^\ast_{n+1}, {\cal A} \} ,
  \end{equation}
with ${\cal Z}_{ij}$ given by
  \begin{equation}
    {\cal Z}_{ij} = \left(
      \bfm{T}^{(2)}_{n+1} - \bfm{T}^{(2)}_n - \mathbbm{E}_{n+1} [
      \bfm{E}^{(2)}_{n+1} - \bfm{E}^{(2)}_n - \Delta\hat{\lambda} \hat{\bfm{N}}{}^\ast_{n+1} ]
      \right)_{ij} ,
  \end{equation}


Although the solution sought in the plastic state ($F^{\ast\it trial}_{n+1}>0$)
corresponds to a strictly positive $\Delta\lambda$, the above simple treatment leads
to a significant increase of the radius of convergence of the Newton method
(actually, now a semi-smooth Newton method due to the $\max(\cdot)$ function in the
definition of $\Delta\hat{\lambda}$) so that the computations may proceed with
larger time increments.



\bibliographystyle{unsrt}
\bibliography
{%
../BIBTEX/contact,%
../BIBTEX/lubrication,%
../BIBTEX/stupkiewicz,%
../BIBTEX/micromechanics,%
../BIBTEX/phasetrans,%
../BIBTEX/roaz%
}

\begin{thebibliography}{10}

\bibitem{SimoHughes98}
J.~C. Simo and T.~J.~R. Hughes.
\newblock {\em Computational Inelasticity}.
\newblock Springer-Verlag, New York, 1998.

\bibitem{SouzaNeto08}
E.~A. de~Souza~Neto, D.~Peric, and D.~R.~J. Owen.
\newblock {\em Computational Methods for Plasticity: Theory and Applications}.
\newblock Wiley, 2008.

\bibitem{Borja98}
R.~I. Borja and C.~Tamagnini.
\newblock Cam-clay plasticity part iii: Extension of the infinitesimal model to
  include finite strains.
\newblock {\em Comput. Method. Appl. M.}, 155:73--95, 1998.

\bibitem{MeschkeLiu99}
G.~Meschke and W.~N. Liu.
\newblock A re-formulation of the exponential algorithm for finite strain
  plasticity in terms of {C}auchy stresses.
\newblock {\em Comp. Meth. Appl. Mech. Engng.}, 173:167--187, 1999.

\bibitem{Perez03}
A.~Perez-Foguet, A.~Rodriguez-Ferran, and A.~Huerta.
\newblock Efficient and accurate approach for powder compaction problems.
\newblock {\em Comput. Mech.}, 30:220--234, 2003.

\bibitem{Ortiz04}
M.~Ortiz and A.~Pandolfi.
\newblock A variational cam-clay theory of plasticity.
\newblock {\em Comput. Method. Appl. M.}, 193:27--29, 2004.

\bibitem{Rouainia06}
M.~Rouainia and D.~M. Wood.
\newblock Computational aspects in finite strain plasticity analysis of
  geotechnical materials.
\newblock {\em Mech. Res. Commun.}, 33:123--133, 2006.

\bibitem{Frenning07}
G.~Frenning.
\newblock Analysis of pharmaceutical powder compaction using multiplicative
  hyperelasto-plastic theory.
\newblock {\em Powder Technol.}, 172:103--112, 2007.

\bibitem{Karrech11}
A.~Karrech, K.~Regenauer-Lieb, and T.~Poulet.
\newblock Frame indifferent elastoplasticity of frictional materials at finite
  strain.
\newblock {\em Int. J. Solids Struct.}, 48:397--407, 2011.

\bibitem{Piccolroaz06-1}
A.~Piccolroaz, D.~Bigoni, and A.~Gajo.
\newblock An elastoplastic framework for granular materials becoming cohesive
  through mechanical densification. {P}art {I} -- small strain formulation.
\newblock {\em Eur. J. Mech. A/Solids}, 25:334--357, 2006.

\bibitem{Piccolroaz06-2}
A.~Piccolroaz, D.~Bigoni, and A.~Gajo.
\newblock An elastoplastic framework for granular materials becoming cohesive
  through mechanical densification. {P}art {II} -- the formulation of
  elastoplastic coupling at large strain.
\newblock {\em Eur. J. Mech. A/Solids}, 25:358--369, 2006.

\bibitem{StupPicBig14}
S.~Stupkiewicz, A.~Piccolroaz, and D.~Bigoni.
\newblock Elastoplastic coupling to model cold ceramic powder compaction.
\newblock {\em J. Eur. Cer. Soc.}, 2014.
\newblock doi:10.1016/j.jeurceramsoc.2013.11.017.

\bibitem{BigoniPiccolroaz04}
D.~Bigoni and A.~Piccolroaz.
\newblock Yield criteria for quasibrittle and frictional materials.
\newblock {\em Int. J. Sol. Struct.}, 41:2855--2878, 2004.

\bibitem{Korelc02}
J.~Korelc.
\newblock Multi-language and multi-environment generation of nonlinear finite
  element codes.
\newblock {\em Engineering with Computers}, 18:312--327, 2002.

\bibitem{Korelc09}
J.~Korelc.
\newblock Automation of primal and sensitivity analysis of transient coupled
  problems.
\newblock {\em Comp. Mech.}, 44:631--649, 2009.

\bibitem{Bigoni00}
D.~Bigoni.
\newblock Bifurcation and instability of non-associative elastoplastic solids.
\newblock In H.~Petryk, editor, {\em Material Instabilities in Elastic and
  Plastic Solids}, volume 414 of {\em CISM Courses and Lectures}, pages 1--52.
  Springer-Verlag Wien, 2000.

\bibitem{StupDenPicBig14:sub}
S.~Stupkiewicz, R.~P. Denzer, A.~Piccolroaz, and D.~Bigoni.
\newblock Implicit yield function formulation for granular and rock-like
  materials.
\newblock (submitted).

\bibitem{HillRice73}
R.~Hill and J.~R. Rice.
\newblock Elastic potentials and the structure of inelastic constitutive laws.
\newblock {\em SIAM J. Appl. Math.}, 25:448--461, 1973.

\bibitem{BrannonLeelavanichkul10}
R.~M. Brannon and S.~Leelavanichkul.
\newblock A multi-stage return algorithm for solving the classical damage
  component of constitutive models for rocks, ceramics, and other rock-like
  media.
\newblock {\em Int. J. Fract.}, 163:133--149, 2010.

\bibitem{Penasa14:sub}
M.~Penasa, A.~Piccolroaz, L.~Argani, and D.~Bigoni.
\newblock Integration algorithms of elastoplasticity for ceramic powder
  compaction.
\newblock {\em J. Eur. Cer. Soc.}
\newblock doi:10.1016/j.jeurceramsoc.2014.01.041.

\bibitem{WeberAnand90}
G.~Weber and L.~Anand.
\newblock Finite deformation constitutive equations and a time integration
  procedure for isotropic, hyperelastic--viscoplastic solids.
\newblock {\em Comp. Meth. Appl. Mech. Engng.}, 79:173--202, 1990.

\bibitem{SteinmannStein96}
P.~Steinmann and E.~Stein.
\newblock On the numerical treatment and analysis of finite deformation ductile
  single crystal plasticity.
\newblock {\em Comp. Meth. Appl. Mech. Engng.}, 129:235--254, 1996.

\bibitem{Michaleris94}
P.~Michaleris, D.~A. Tortorelli, and C.~A. Vidal.
\newblock Tangent operators and design sensitivity formulations for transient
  non-linear coupled problems with applications to elastoplasticity.
\newblock {\em Int. J. Num. Meth. Engng.}, 37:2471--2499, 1994.

\bibitem{PerezFoguetArmero02}
A.~Perez-Foguet and F.~Armero.
\newblock On the formulation of the closest-point projection algorithms in
  elastoplasticity---part {II}: {G}lobally convergent schemes.
\newblock {\em Int. J. Num. Meth. Engng.}, 53:331--374, 2002.

\bibitem{KorelcStupkiewicz14}
J.~Korelc and S.~Stupkiewicz.
\newblock Closed-form matrix exponential and its application in finite-strain
  plasticity.
\newblock {\em Int. J. Num. Meth. Engng.}, 2014.
\newblock doi:10.1002/nme.4653.

\bibitem{KorelcWriggers97}
J.~Korelc and P.~Wriggers.
\newblock Improved enhanced strain four-node element with {T}aylor expansion of
  the shape functions.
\newblock {\em Int. J. Num. Meth. Engng.}, 40:407--421, 1997.

\bibitem{AlartCurnier91}
P.~Alart and A.~Curnier.
\newblock A mixed formulation for frictional contact problems prone to {N}ewton
  like solution methods.
\newblock {\em Comp. Meth. Appl. Mech. Engng.}, 92:353--375, 1991.

\bibitem{LengKorStu11}
J.~Lengiewicz, J.~Korelc, and S.~Stupkiewicz.
\newblock Automation of finite element formulations for large deformation
  contact problems.
\newblock {\em Int. J. Num. Meth. Engng.}, 85:1252--1279, 2011.

\bibitem{Bazant98}
Z.~P. Bazant.
\newblock Easy-to-compute tensors with symmetric inverse approximating {H}encky
  finite strain and its rate.
\newblock {\em Trans. ASME J. Eng. Mat. Technol.}, 120:131--136, 1998.

\end{thebibliography}

\end{document}